\documentclass[12pt]{article}
\usepackage{epsf}
\def\beq{\begin{equation}}
\def\eeq{\end{equation}}
\def\ap#1#2#3 {Ann. Phys. (NY) {\bf#1} (19#2) #3}
\def\err#1#2#3 {{\it Erratum} {\bf#1} (19#2) #3}
\def\ib#1#2#3 {{\it ibid.} {\bf#1} (19#2) #3}
\def\ijmp#1#2#3 {Int. J. Mod. Phys. {\bf#1} (19#2) #3}
\def\jetp#1#2#3 {JETP Lett. {\bf#1} (19#2) #3}
\def\mpl#1#2#3 {Mod. Phys. Lett. {\bf#1} (19#2) #3}
\def\np#1#2#3 {Nucl. Phys. {\bf#1} (19#2) #3}
\def\pl#1#2#3 {Phys. Lett. {\bf#1} (19#2) #3}
\def\prep#1#2#3 {Phys. Rep. {\bf#1} (19#2) #3}
\def\prev#1#2#3 {Phys. Rev. {\bf#1} (19#2) #3}
\def\prl#1#2#3 {Phys. Rev. Lett. {\bf#1} (19#2) #3}
\def\sjnp#1#2#3 {Sov. J. Nucl. Phys. {\bf#1} (19#2) #3}
\def\spj#1#2#3 {Sov. Phys. JETP {\bf#1} (19#2) #3}
\def\spu#1#2#3 {Sov. Phys. Usp. {\bf#1} (19#2) #3}
\def\zp#1#2#3 {Zeit. Phys. {\bf#1} (19#2) #3}
\def\a{\alpha}
\def\drcn{\delta R^{c/n}}

\begin{document}
\begin{titlepage}
\begin{center}
{\Large \bf Theoretical Physics Institute \\
University of Minnesota \\}  \end{center}
\vspace{0.2in}
\begin{flushright}
TPI-MINN-03/01-T \\
UMN-TH-2125-03 \\
January 2003 \\
\end{flushright}
\vspace{0.3in}
\begin{center}
{\Large \bf  Variation of the relative yield of  charged and neutral $B$
mesons across the $\Upsilon(4S)$ resonance
\\}
\vspace{0.2in}
{\bf M.B. Voloshin  \\ }
Theoretical Physics Institute, University of Minnesota, Minneapolis,
MN
55455 \\ and \\
Institute of Theoretical and Experimental Physics, Moscow, 117259
\\[0.2in]
\end{center}

\begin{abstract}
It is shown that the ratio of the production rates of the pairs $B^+B^-$
and $B^0 {\overline B^0}$ should experience a substantial and rapid
variation with energy within the width of the $\Upsilon(4S)$ resonance,
crossing the value of one near the center of the resonance. This
behavior is due to an interference of the rapidly changing with energy
Breit-Wigner phase with the phase introduced in the wave function of
charged mesons by their Coulomb interaction.
\end{abstract}

\end{titlepage}

The ratio of the production rates of charged and neutral $B$ mesons in
$e^+e^-$ annihilation at the $\Upsilon(4S)$ resonance,
\beq
R^{c/n}=1+\drcn={\sigma(e^+e^- \to B^+B^-) \over \sigma(e^+e^- \to B^0
{\overline B^0})}~,
\label{rcn}
\eeq
is an important parameter in detailed studies of the properties of $B$
mesons. Recent dedicated measurements\cite{cleo1,babar,cleo2} of
$R^{c/n}$ at the maximum of the resonance report values ranging from
$1.04\pm 0.07 \pm0.04$ \cite{cleo1} to $1.10 \pm 0.06 \pm 0.05$
\cite{babar}, which leave enough room for further studies of the
quantity of interest $\drcn$.

Theoretically the difference $\drcn$ of the discussed ratio from one
arises as dominantly an effect of the Coulomb interaction, clearly
different for charged and neutral $B$ mesons, since the mass difference
$m_{B^0}-m_{B^+}=0.33 \pm 0.28 \, MeV$\cite{pdg} is quite small, and its
effect can be accounted separately. In the most simple
approach\cite{am}, where the $B$ mesons are treated as point particles,
and the existence of the resonant interaction is ignored, the estimate
of $\drcn$ can be expressed in terms of the c.m. velocity $v$ of
produced $B$ mesons, using the textbook Coulomb wave functions:
\beq
\drcn = {\pi \a \over 2 \, v} + O\left ({\a^2 \over v^2} \right )~.
\label{drcn0}
\eeq
At the excitation energy of the $\Upsilon(4S)$ resonance, $E_0
=M_{\Upsilon(4S)}-2 m_B \approx 20 \, MeV$,  one has $v \approx 0.06$,
and the simple estimate (\ref{drcn0}) would yield $\drcn \approx 0.19$.
It was subsequently argued\cite{lepage} that the estimate of $\drcn$ can
in fact be substantially reduced from eq.(\ref{drcn0}) if one accounts
for a finite size of the $B$ mesons (through their electromagnetic form
factor) and also for the finite size of the $\Upsilon(4S)$. The latter
effect was further discussed in a specific model of heavy
quarkonium\cite{be}. Recently the problem of calculation of the ratio
$R^{c/n}$ was revisited\cite{kmm} in the context of a chiral-type model
for strong interaction of $B$ mesons at short distances, including the
$B^*B\pi$ vertex and the coupled channels with pairs of pseudoscalar
and/or vector mesons, although still considering all the mesons as
point-like with respect to the Coulomb interaction.

It should be noted that in all previous theoretical studies of the ratio
$R^{c/n}$ the presence of the resonance in the wave function of the $B$
meson pair was essentially ignored. In other words, the coupling of the
resonance to the $B$ mesons was either treated perturbatively (although
with a form factor\cite{lepage,be}), or the considered model of the
strong interaction did not contain a resonance at all\cite{kmm}. For
this reason the results predicted a smooth behavior of $\drcn$ with
energy in the region of the $\Upsilon(4S)$ resonance.  It is however
well known (see e.g. in the two last chapters of the textbook \cite{ll})
that the presence of a resonance produces a large effect on the wave
function of the scattering states, which is rapidly changing across the
resonance with energy. The scale for the variation is set by the
resonance width $\Gamma$. In particular, the relative phase $2 \delta$
between the outgoing and incoming spherical waves (twice the scattering
phase $\delta$) changes by $2 \pi$ at the scale $\Gamma$ when the
excitation energy $E$ passes the central value $E_0$. It is the purpose
of the present paper to properly take into account the resonant behavior
of the wave function of the scattering states along the lines of the
standard non relativistic scattering theory\cite{ll}. It will be shown
that an interplay between the rapidly changing relative phase of
incoming/outgoing wave  and of the effects of the Coulomb interaction
gives rise to a rather non-trivial behavior of $\drcn$ with the energy
changing across the resonance. Namely $\drcn$ has to change sign  at
energy within a fraction of the width $\Gamma$ from the `nominal'
resonance center energy $E_0$. It will also be argued that the effect of
a rapid variation of $\drcn$ should be model independent, while the
details, such as the overall magnitude of $\drcn$ and the precise
position of its zero(s), do depend on yet unknown details of the strong
and electromagnetic interactions of the $B$ mesons at short distances
and of the structure of the $\Upsilon(4S)$ resonance. Thus a detailed
experimental study of the behavior of $\drcn$ in the resonance region
could in principle provide a certain insight into those finer properties
of the hadron dynamics.

The standard physical picture for considering the scattering in the
resonance region (c.f. Ref.\cite{ll}) is that the strong interaction,
responsible for the existence of the resonance has a short range $a$,
and the essential effects of the interaction at distances $r < a$ can be
parameterized in terms of phenomenologically measurable parameters of
the resonance, the most important being its energy $E_0$ above the
threshold and the width $\Gamma$. At distances larger than $a$ the
motion is described by a known potential $V(r)$: either $V(r)=0$, or a
Coulomb potential (with a possible modification due to form factor at
short distances), where the wave function of the scattering state can be
found explicitly from the Schr\"odinger equation. The boundary
(matching) conditions at $r \approx a$ for the `outer' wave function are
related to the measurable parameters of the resonance. In the discussed
process the $B \overline B$ pairs are produced in the $P$ wave. Also
beyond the region of strong interaction, i.e. at $r > a$, there is no
strong interaction mixing between the ``neutral", $B^0 \overline B^0$,
and the ``charged", $B^+ B^-$, channels. Thus the `outer' wave function
at $r > a$ is described by the spherical wave with $L=1$, whose radial
part can be written as $R(r)=\chi(r)/r$, with a separate function
$\chi(r)$ for each of the channels: $\chi_n(r)$ and $\chi_c(r)$, each
satisfying at energy $E=p^2/m$ the corresponding one-dimensional
Schr\"odinger equation
\beq
\chi_n'' + \left ( p^2 - {2 \over r^2} \right ) \, \chi_n
=0~,~~~~\chi_c'' + \left ( p^2 + m \, {\a \over r} - {2 \over r^2}
\right ) \, \chi_c =0~,
\label{schr}
\eeq
where the prime denotes derivative over $r$, and $m=m_B \approx 5280\,
MeV$ is the mass of either of the $B$ mesons (a possible small mass
difference between the charged and neutral $B$ mesons is completely
ignored throughout the present discussion).

The coupling between the ``neutral" and the ``charged" channels takes
place in the region of strong interaction at short distances. At those
distances the light quark parts of the mesons strongly overlap and
become a part of (presumably) quite complicated dynamics of light quarks
and gluons. Thus in this region it would be inappropriate to continue
description in terms of individual $B$ mesons. The boundary condition
for the `outer' dynamics at distances $r > a$ is however dictated by the
isotopic invariance of the strong interaction. Namely, one can assume
with a rather high degree of accuracy that when the $B$ mesons emerge
from the region of strong dynamics as individual particles their wave
function is an isotopic singlet. In other words, the isospin condition
for the functions $\chi_n$ and $\chi_c$ is that they evolve from one and
the same function at a certain short distance $r=a$, i.e. that
\beq
\chi_c(a)=\chi_n(a)~~~{\rm and}~~~\chi_c'(a)=\chi_n'(a)~,
\label{bc}
\eeq
which boundary conditions can be viewed as our formal definition of the
short distance parameter $a$. Although there can be a small `intrinsic'
isospin  violation also in the region of strong interaction, its effect
in $\drcn$, as discussed in Ref.\cite{kmm}, is noticeably smaller than
that of the Coulomb interaction, and can be studied as a further
adjustment, using the approximation of exact isospin symmetry at short
distances in eq.(\ref{bc}) as a starting point.

As is known from the standard Breit-Wigner description of a resonance
scattering\cite{ll} at energy $E$ near the position $E_0$ of the
resonance, the relevant `outer' solution of the Schr\"odinger equations
(\ref{schr}) for stationary wave functions has the form
\begin{eqnarray}
\chi_n(r)&=&(\Delta-i \, \gamma)\, b_n \, f_n(r)+ (\Delta+i \, \gamma)\,
b_n^* \, f_n^*(r)~, \nonumber \\
\chi_c(r)&=&(\Delta-i \, \gamma)\, b_c \, f_c(r)+ (\Delta+i \, \gamma)\,
b_c^* \, f_c^*(r)~,
\label{wfs}
\end{eqnarray}
where $\Delta=E-E_0$, $\gamma=\Gamma/2$ and the complex coefficients
$b_{n(c)}$ are generally functions of the energy, which however have no
zeros at $\Delta=i \, \gamma$. Finally, each of the functions $f_n$ and
$f_c$ is the solution of the corresponding equation in (\ref{schr}),
which contains only the outgoing wave, i.e. at $r \to \infty$ they
contain only the factor $\exp (i p r)$ (while their complex conjugates
$f^*_{n(c)}$ contain only the incoming wave factor $\exp (-i p r)$).

The function $f_n(r)$ specified by this condition is well known for the
free motion with $L=1$,
\beq
f_n(r)=\left ( 1+ {i \over p r} \right ) e^{ipr}~,
\label{ffree}
\eeq
and with this condition for its phase, the phase of the coefficient
$b_n$ coincides with the non-resonant scattering phase $\delta_1$ at
$L=1$,
\beq
\exp(2i \delta_1)= {b_n \over b_n^*}~.
\label{d1}
\eeq
The corresponding function $f_c(r)$ for the motion in the Coulomb
potential is also well known (see e.g. in Ref.\cite{ll}), however for
our present purpose it would be more convenient to make use of the
perturbation theory in the Coulomb interaction, rather than to do an
expansion of the explicit expression. In specifying the phase convention
for the function $f_c$ a minor technical point arises due to the well
known fact that its phase at $r \to \infty$ contains a slowly varying
logarithmic Coulomb phase:
$f_c(r) \sim \exp i [pr + (\a/2v)\, \ln 2pr + const]$. This however can
be readily resolved by assuming that the Coulomb interaction is cut off
at a large distance $r = R$. Then at still larger $r$ the Coulomb phase
does not change and can be considered as constant. Clearly the physical
results, including the discussed here effect in $\drcn$ do not depend on
this infrared cutoff. Implying such regularization, the function
$f_c(r)$ in eq.(\ref{wfs}) can be chosen to exactly coincide (both in
phase and in normalization) with $f_n(r)$ at asymptotically large
distances: $f_c(r) = \exp(ipr)$. Thus any difference in phase and
magnitude that arises from the Coulomb interaction in the ``charged"
channel is encoded in the coefficient $b_c$.

In order to find from the wave function of a stationary state
(eq.(\ref{wfs})) the relative rate of production of the pairs of charged
and neutral $B$ mesons in $e^+e^-$ annihilation, i.e. by a source
localized well inside the region of strong interaction, it is necessary
to note that the rate in each channel is proportional to the {\it
inverse} of the norm squared of the coefficient in front of the {\it
incoming} wave. This can be understood by considering the reverse
process: annihilation of a meson pair into $e^+e^-$, in complete analogy
with an explanation of the ``$|\psi(0)|^2$ rule" for production of bound
states. In the reverse process the incoming wave has a fixed flux, i.e.
a fixed norm of the coefficient in front of $\exp(-i p r)$ at large $r$.
Matching this wave to the incoming part of the wave function in
eq.(\ref{wfs}) implies that the corresponding function $\chi(r)$ has to
be divided by the coefficient of its incoming wave part. Under the
normalization conventions adopted here for the functions $f(r)$ this
results in the annihilation rate being proportional to the factor
$|(\Delta+i \, \gamma)\,b^*|^{-2}$, with $b$ equal to $b_n$ or $b_c$,
depending on the chosen initial state for the incoming wave. Clearly,
this factor in the rate contains both the standard Breit-Wigner
resonance curve and the normalization factor $|b|^{-2}$. Thus the
discussed here ratio of the yields in the two channels is given
by\footnote{In a somewhat more widely familiar non-resonant situation
when point-like particles are produced by a point source, the functions
$\chi_n(r)$ and $\chi_c(r)$ are the regular at $r=0$ solutions of the
equations (\ref{schr}). In this case one has (for a $P$ wave): $|b_n /
b_c|^2 = |\psi'_c(0)/\psi'_n(0)|^2$, where $\psi_c(r)$ and $\psi_n(r)$
are the wave functions for corresponding stationary states, having the
same relative normalization at infinity.}
\beq
R^{c/n}={|b_n|^2 \over |b_c|^2}~.
\label{rb}
\eeq

The relation between the coefficients $b_n$ and $b_c$ is found from the
matching conditions (\ref{bc}).
After substituting the wave functions from eq.(\ref{wfs}) into the
conditions (\ref{bc}), the ratio of the coefficients (in fact the
inverse of that entering eq.(\ref{rb})) is found as
\begin{eqnarray}
{b_c \over b_n}&=&{{f_c'}^*\left ( f_n  + f_n^* \,e^{-2i \delta_{BW} -
2i \delta_1} \right ) - f_c^* \left ( f_n'  + {f_n'}^*\, e^{-2i
\delta_{BW} - 2i \delta_1} \right ) \over {f_c'}^* f_c - f_c^*
f_c'}\nonumber \\
&=&{i \over 2 p} \left [ {f_c'}^*\left ( f_n  + f_n^* \,e^{-2i
\delta_{BW} - 2 i \delta_1} \right ) - f_c^* \left ( f_n'  + {f_n'}^*\,
e^{-2i \delta_{BW} - 2i \delta_1} \right ) \right ]~,
\label{genf}
\end{eqnarray}
where all the functions and their derivatives are taken at the matching
point $r=a$, $\delta_{BW}$ is the standard Breit-Wigner resonance phase:
$\exp (2 i \delta_{BW})=(\Delta - i \, \gamma)/(\Delta + i \, \gamma)$,
and the non-resonant phase $\delta_1$ is defined by eq.(\ref{d1}).
Finally, in the last transition a use is made of the fact that the
denominator in the intermediate expression is the Wronskian, which is
constant in $r$ and can thus be found from the asymptotic form of the
function $f_c$ at large $r$.

The equation (\ref{genf}) contains no approximation with regards to the
Coulomb interaction, and can be used down to arbitrarily small values of
$v$, i.e. for arbitrary values of the Coulomb parameter $\a/v$, provided
that the exact Coulomb function is used for $f_c(r)$. However for the
practical purpose of discussing the Coulomb effects at the
$\Upsilon(4S)$ resonance it is sufficient to consider only the effect of
first order in $\a$. In this order one can write $f_c(r)=f_n(r)+\phi(r)$
with $\phi$ being formally a small perturbation of order $\a$ of the
wave function. Using this expression in eq.(\ref{genf}) and also
assuming an expansion of the ratio $b_c/b_n$ to the first order in $\a$,
one readily finds
\beq
\drcn={|b_n|^2 \over |b_c|^2}-1= {1 \over p}\, {\rm Im}\left \{ e^{2i
\delta_{BW} + 2i \delta_1} \, \left[ \phi(a) \, f'_n(a)-\phi'(a) \,
f_n(a) \right ]    \right \}~.
\label{linf}
\eeq
The combination of the functions in this expression can be found
directly from the equations (\ref{schr}) by also a rather standard
method. Indeed, the function $f_n(r)$ satisfies the first of those
equations, while $f_c(r)$ satisfies the second equation. Multiplying the
first equation by $f_c(r)$ and the second equation by $f_n(r)$,
subtracting the results, and expanding in the difference $\phi$ between
$f_c$ and $f_n$, one arrives at the relation
\beq
{d \over dr} \, \left [ \phi(r) \, f_n'(r)  - \phi'(r)\, f_n(r) \right ]
= m \, {\a \over r} \, f_n^2(r)~.
\eeq
Integrating this relation over $r$ from $a$ to infinity, using the fact
that under our conventions $\phi(r) \to 0$ at $r \to \infty$, and also
using the explicit expression (\ref{ffree}) for $f_n(r)$, one finds in
terms of eq.(\ref{linf}) the formula
\begin{eqnarray}
\label{final}
\drcn &=& -{\a \over v} \, {\rm Im}\left [ e^{2i \delta_{BW} + 2i
\delta_1} \, \int_a^\infty e^{2ipr} \, \left ( 1+ {i \over p r} \right
)^2 \, {dr \over r} \right ]  \\
&=& {\a \over v} \, \left [ {\Delta^2 - \gamma^2 \over \Delta^2 +
\gamma^2}\, \left ( A \, \cos 2 \delta_1 + B \, \sin 2 \delta_1 \right )
-  {2 \, \gamma \, \Delta \over \Delta^2 + \gamma^2} \, \left ( B\, \cos
2 \delta_1 - A \, \sin 2 \delta_1 \right ) \right ]~,\nonumber
\end{eqnarray}
where in the latter expression the coefficients $A$ and $B$ are given
(with the oposite sign) by respectively the imaginary and the real part
of the integral with complex exponent:
\begin{eqnarray}
A &=& -\int_{pa}^\infty \left [  \left (1-{1 \over u^2} \right )\,\sin
2u  + {2 \, \cos 2u \over u} \right ] \, {du \over u}~,\nonumber \\
B &=& \int_{pa}^\infty \left [ {2 \,\sin 2 u \over u}- \left (1- {1
\over
u^2} \right ) \, \cos 2 u  \right ] \, {du \over u}~.
\label{ab}
\end{eqnarray}
At small values of the product $pa$ the coefficients $A$ and $B$ have
the expansion:
\beq
A={\pi \over 2} - {2 \, p \, a \over 3}+ O(p^3 a^3)~,~~~B={1 \over 2\,
p^2 \, a^2} - \ln 2 p a - \gamma_E +1 + O(p^4 a^4)~,
\label{spa}
\eeq
where $\gamma_E=0.577\ldots$ is the Euler constant.

One can see from eq.(\ref{final}) that in the limit, where the effects
of the strong interaction in the wave function of the scattering state
are ignored, corresponding to $\gamma \to 0$ and $\delta_1 \to 0$, the
simple estimate (\ref{drcn0}) of $\drcn$ is recovered, assuming
production of point-like particles by a point source (i.e. also $a \to
0$).
However this limit of a vanishing resonance width is totally inadequate
for resolving the behavior of $\drcn$ inside the resonance curve, i.e.
at the energy scale of $\gamma$. In particular, the second term in the
final expression in eq.(\ref{final}), proportional to $\gamma \Delta
/(\Delta^2 + \gamma^2)$ vanishes at energies far away from the
resonance, i.e.  at $|\Delta| \gg \gamma$, as well as at the center of
the resonance, i.e. at $\Delta=0$. On the other hand, the first term
contains the factor $(\Delta^2 - \gamma^2)/(\Delta^2 + \gamma^2)$, which
changes from $+1$ away from the resonance to $-1$ at $\Delta=0$. Thus
$\drcn$ necessarily has to change sign within the width of the
resonance.

Given that the coefficient $B$ is singular at small momenta, it is
instructive to analyze the behavior of $\drcn$ described by the
equations (\ref{final} - \ref{spa}) in the limit of small $p$ (but still
considering $\a /v$ as small for the applicability of the perturbative
treatment of the Coulomb interaction).  At this point one has to recall
that the width parameter $\gamma$ for a $P$ wave resonance has to vanish
at small $p$ as $p^3$, and that the non-resonant scattering phase
$\delta_1$ also vanishes as $p^3$. Thus the effect of the $p^{-2}$
singularity in the coefficient $B$ results in a constant at small
velocity term in the correction: $const \cdot \a$, which is small in
comparison with the dominant part of the correction behaving as $\a/v$.

It should be emphasized, prior to discussing the behavior at the
realistic $\Upsilon(4S)$ resonance, that the expressions (\ref{final}) -
(\ref{spa}) are formally applicable only in the limit of small $pa$.
Indeed, in this limit the details of the transition between an
isotopically symmetric strong dynamics at short disatnces and the
Coulomb behavior at the relevant distances of order $p^{-1}$ are not
essential, and the parameter $a$ enters only the leading singularity of
the coefficient $B$ resulting in a subleading at low energy term in
$\drcn$, which can be studied phenomenologically\footnote{It can be
noticed that a non-removable dependence on a short-distance parameter is
a general property of scattering in the states with $L \neq 0$, and in
fact it also appears, although as a logarithmic dependence, in the $S$
wave scattering in the presence of a Coulomb interaction\cite{ll}.}.
When the parameter $pa$ cannot be considered as small, the details of
the actual behavior of the wave functions at
the transition distances would generally depend on the shape of the
transition, and a more elaborate matching at short distances, than the
conditions (\ref{bc}) at a fixed distance,  might be required.

The position of the actual $\Upsilon(4S)$ resonance corresponds to the
momentum $p_0 \approx 330 MeV$. Thus, most likely, the relevant values
of the parameter $p a$ are of order one, i.e. the parameter is neither
small nor large in the region of interest. Under these circumstances one
can approach this region of phenomenological interest from the side of
small values of $pa$ aiming at an at least qualitative description of
what kind of behavior should be expected for $\drcn$ at the resonance.
In addition, the width of the $\Upsilon(4S)$, $\Gamma \approx 14\, MeV$,
is not small as compared to its excitation energy. Combined with the not
too small value of $p_0$ this can generally lead to that effects
of higher terms of expansion in $p^2$ of the quantities $\gamma(p)$,
$\delta_1(p)$, and $\Delta(p)$ may become essential. (An attempt at
taking into account higher than $p^3$ terms in the width parameter was
done in connection with the experimental measurement of $\Gamma$ in
Ref.\cite{argus}. However the effect of those higher terms turned out to
be quite small within the experimental accuracy.)
\begin{figure}[ht]
  \begin{center}
    \leavevmode
    \epsfxsize=12cm
    \epsfbox{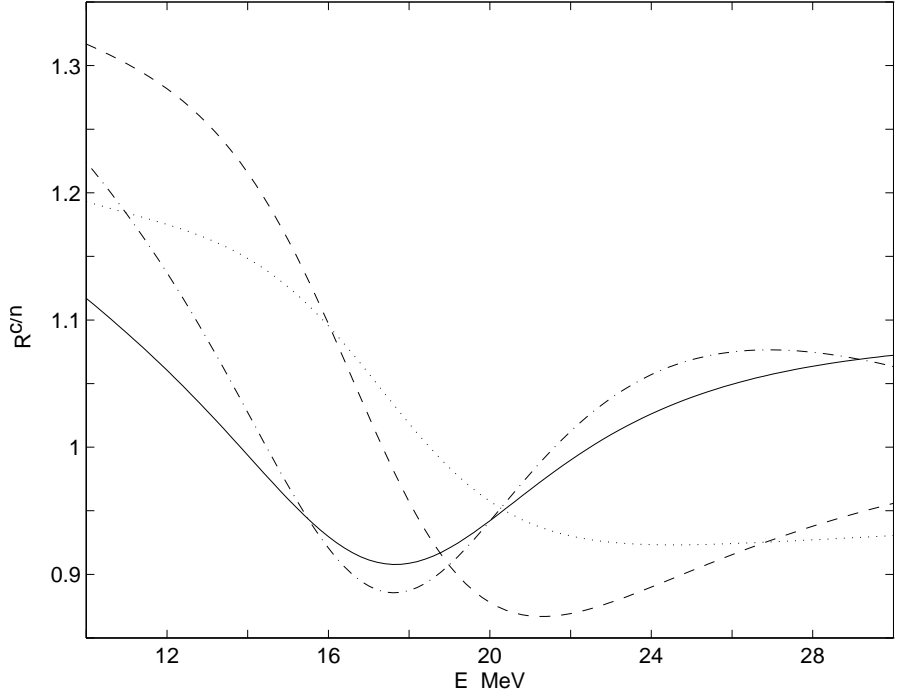}
    \caption{The dependence of the ratio $R^{c/n}$ on the excitation
energy $E=\sqrt{s}-2 m_B$ in the region of the $\Upsilon(4S)$ resonance
(the center position is assumed to be at $E_0=20 \, MeV$) for some
values of $a$ and $\delta_1(E_0)$: $a^{-1}=200 \, MeV,~\delta_1(E_0)=0$
(solid), $a^{-1}=400 \, MeV,~\delta_1(E_0)=0$ (dashed),
$a^{-1}=300 \, MeV,~\delta_1(E_0)=30^0$ (dashdot), and $a^{-1}=300 \,
MeV,~\delta_1(E_0)=-30^0$ (dotted).}
  \end{center}
\label{fig:xy}
\end{figure}

With all the stated reservations about uncertainties involved in a
quantitative description of the behavior of $\drcn$ in the region of the
$\Upsilon(4S)$ resonance, a qualitative illustration of the expected
variation of $R^{c/n}$ in the resonance region is provided by the plots
in Fig.1. The curves in the plots are calculated with various  values of
$a$ and $\delta_1$ under the following assumptions: only the leading
terms in the expansion of $A$ and $B$ at small $pa$, explicitly shown in
eq.(\ref{spa}), are retained, the width parameter is parameterized as
$\gamma=(\Gamma/2)\,(p/p_0)^3$, and the non-resonant phase as
$\delta_1=\delta_1(E_0) \, (p/p_0)^3$. (It should be mentioned that the
range of $\delta_1(E_0)$ from $-30^0$ to $+30^0$ most likely is
unrealistically broad, and is used here for an illustration of the
effect of the phase under extreme assumptions.) One can clearly see from
the shown curves that being undoubtedly different in details, they
exhibit quite similar qualitative behavior\footnote{The second zero of
$\drcn$ at $E > E_0$ is not reached in some of the curves in Fig.1
within the shown range of the energy. The behavior at large energies
however becomes very sensitive to higher in $p^2$ terms in $\gamma$,
$\Delta$, and $\delta_1$, and the location of that zero cannot be
presently estimated with any certainty.}. Thus, as expected on general
grounds from eq.(\ref{final}), the very fact of a substantial and rapid
variation of $\drcn$ within the resonance width stays robust under
assumptions about the presently unknown parameters.

The strong variation of $\drcn$ at the scale of few MeV near the center
of the $\Upsilon(4S)$ resonance can be important in a comparison of the
data obtained in different experiments, especially at different
electron-positron colliders, given the differences in the beam energy
spread and possible slight differences in the values (and stability) of
the central energy of the beams, at which the data are collected. The
thus far available results\cite{cleo1,babar,cleo2} are consistent with
each other mainly due to the substantial statistical and systematic
errors. This agreement however can change if the measurements of the
discussed relative yield are pursued further with better accuracy,
possibly permitting a quantitative study of the variation of this yield.
As is argued here, the effect of the variation is not small, and the
detailed behavior is sensitive to properties of the heavy and light
quark and hadron dynamics, which would be difficult, if possible at all,
to study by other means.

I gratefully acknowledge useful discussions with Ron Poling, Jon Urheim,
and Arkady Vainshtein of the experimental and theoretical issues related
to the problem considered in this paper. This work is supported in part
by the DOE grant DE-FG02-94ER40823.

\end{document}